# Observation of Moiré Plasmonic Skyrmion Clusters

Lan Zhang[1], Lipeng Wan[1,2,*], Weimin Deng[1,*], Liang Hou[1], Jumin Qiu[1],
Qiushun Zou[3], Tongbiao Wang[1], Daomu Zhao[4], Tianbao Yu[1,*]

[1] School of Physics and Material Science & Jiangxi Provincial Key Laboratory of Photodetectors, Nanchang University, Nanchang 330031, China;
[2] Leiden Institute of Physics, Leiden University, Leiden, CA 2333, The Netherlands;
[3] Faculty of Electrical Engineering and Computer Science, Laboratory of Infrared Material and Devices, Zhejiang Key Laboratory of Advanced Optical Functional Materials and Devices, Advanced Technology Research Institute, Ningbo University, Zhejiang 315211, China;
[4] Zhejiang Key Laboratory of Micro-nano Quantum Chips and Quantum Control, School of Physics, Zhejiang University, Hangzhou 310058, China;
* Corresponding Authors: Lwanoptics@ncu.edu.cn; dengweimin@ncu.edu.cn; yutianbao@ncu.edu.cn;

**Skyrmions are topological defects belonging to nontrivial homotopy classes in particle theory. Their remarkably stable topology has recently been observed in electromagnetic waves. For the evanescent fields near a surface, this has been realized so far only for elementary optical skyrmions, with a fixed skyrmion number. Here we report, both in theory and experiment, the concept of moiré plasmonic skyrmion clusters, where multi-skyrmions are nested to form a large optical skyrmion cluster. By leveraging twistronics engineering of plasmonic nanostructures, we demonstrate both crystallized and quasi-crystallized optical skyrmion lattices, revealing an unprecedented degree of topological control. In a misaligned composite nanostructure, the rapid inverting of optical skyrmion number is achieved, which is explained by a lattice model. This topological change of moiré plasmonic skyrmion clusters can serve as a precise beacon of the relative alignment deviation between composite nanostructures.**

## INTRODUCTION

Topology continues to reform optical physics, offering vast opportunities for fundamental research and technological applications, including high-dimensional vortices [1], polarization singularities [2], and correlation switch [3]. Skyrmions are topologically stable quasiparticles, originally proposed as a topological soliton solution by the British particle physicist Tony Skyrme in his study of unified field theory. Apart from the field of high energy physics, skyrmions have been discovered in many other fields, such as Bose–Einstein condensates [4], liquid crystals [5], and magnetic materials [6]. In 2018, two pioneering schemes for realizing optical skyrmion lattices were proposed [7,8], by either utilizing surface plasmon polaritons (SPPs) interference to form on-chip optical polarization skyrmions, or exploiting the topological defects in the local optical spin field, forming an isolated optical spin skyrmion. The time dynamics of optical skyrmion were also well-resolved [9,10]. The concept of skyrmion was later extended to various wave configuration [11–15], such as Stokes parameter of paraxial propagating light [16], pseudo-spin vector in photonic crystals [17], supertoroidal pulses [18], electromagnetic vector of spoof localized surface plasmons [19–21] and optical skyrmion beam generators [22]. Characterized by their ultracompact size and high stability, photonic skyrmions enable the realization of pico-metric displacement sensing [23].

Optical skyrmions are typically generated as isolated unit or periodic lattices with fixed symmetry and skyrmion number. This limits the capacity of photonic devices by encoding information in the topological degree of skyrmions. To advance higher density information storage technologies and photonic devices, high-degree optical skyrmions with tunable skyrmion numbers are essential. Only quite recently, outstanding work has revealed that the symmetry of the photonic skyrmion lattice can be modified by tuning the wavelength [24], but no study has directly been explored in the high-degree optical skyrmion. The major challenge is the fact that the multiple 3D topology structures should be manipulated on a 2D plane simultaneously.

In this work, we theoretically propose and experimentally observe the plasmonic skyrmion clusters, in which multiple Néel-type plasmonic skyrmions nested to form a large optical skyrmion cluster, where the information encoded by the number of skyrmion units exceeding one provides higher storage density for photonic devices. This is achieved using twistronics engineering of plasmonic nanostructure system. The composite six-fold symmetric nanostructure with an engineered twist is used to excites 3D polarization "quasi-particles states" interference on a 2D plane. We found that in the composite nanostructures, the optical skyrmion number is rapidly reversed for the certain alignment deviations. Twisting the composite six-fold symmetric nanostructures enables the generation of moiré optical skyrmions superlattice, which is induced by the interference of dual groups of skyrmion lattices in one plane. This resembles the twisted bilayer graphene [25–27] and twisted bilayer photonic crystals [28–31]. Our proposed skyrmion superlattices may have different crystal or quasicrystal structure with well-defined symmetry and periodicity, depending on the twist angle. Since optical skyrmions are topologically stable quasiparticles, the changes in the twist angle of a few degrees allow us to track the optical skyrmion on the nanoscale. Continuous changes in the twist angle result in the observation of the nucleation as well as the collapse of optical skyrmions.

## RESULTS
### Moiré-induced optical skyrmion clusters

Figure 1 shows the concept and generation method of moiré optical skyrmion clusters. We proposed a twisted plasmonic platform [Fig. 1A], where composite nanostructures with a twist are introduced into the same film. This excites the superposition of well-defined transverse-magnetic (TM) guided waves, leveraging two hexagonal nanoslits with controlled twist angle to generate superimposed plasmonic polarization states. Excited by propagating light with spin angular momentum $\sigma = \pm 1$ impinging on this composite nanostructure, the three-dimensional (3D) polarized electric field on the 2D film is governed by the superposition of dual groups of plasmonic wave, $\mathbf{E}_a(\mathbf{r}, 0)$ and its twisted replica $\mathbf{E}_b(\mathbf{r}, \theta)$. Such surface waves propagate in the transverse plane and evanescently decay in the axial direction. The z-component of the electric field thus takes the form of

$$E^z(\mathbf{r}) = E_a^z(\mathbf{r}, 0) + E_b^z(\mathbf{r}, \theta)$$
$$= e^{-|k_z|z}\left\{\sum_{j=1}^{N=6} \widetilde{E}_j e^{-i(\mathbf{k}_j(0)\cdot\mathbf{r} - \phi_j)} + \sum_{j=1}^{N=6} \widetilde{E}_j e^{-i[\mathbf{k}_j(\theta)\cdot\mathbf{r} - \phi_j - \Delta\phi]}\right\}, \quad (1)$$

where $\mathbf{k}_j(\theta) = \mathbb{R}k_t[\cos\varphi_j, \sin\varphi_j]$ with $\mathbb{R}$ being the 2D unitary rotation matrix $[\cos\theta, -\sin\theta; \sin\theta, \cos\theta]$, $k_t$ and $k_z$ are the transverse and axial components of the wave vector such that $k_t^2 + k_z^2 = k_0^2$, relating to the free-space wavenumber $k_0$. $\widetilde{E}_j$ is the amplitude of each surface wave. $\Delta\phi$ denotes the relative phase difference between the two groups of electric fields, and $\phi_j$ denotes the orientation-associated geometric phase of the $j$th nanoslit, i.e. $\phi_j = 2\pi\sigma j/N$. To ensure the in-phase condition for all nanoslits pairs, two nanoslits is shifted by half an SPP wavelength $\lambda_S$.

According to Eq. (1), we plot the axial field component at $\theta$ = 38.21 degrees. Superimposed optical polarization lattices are observed [Fig. 1B]. For the total interference 3D polarization vector $\bar{\mathbf{E}}$, after normalization, we have

$$\bar{\mathbf{E}}(\mathbf{r}, \theta) = \frac{\text{Re}[\mathbf{E}_a(\mathbf{r},0) + \mathbf{E}_b(\mathbf{r},\theta)]}{|\mathbf{E}_a(\mathbf{r},0) + \mathbf{E}_b(\mathbf{r},\theta)|}. \quad (2)$$

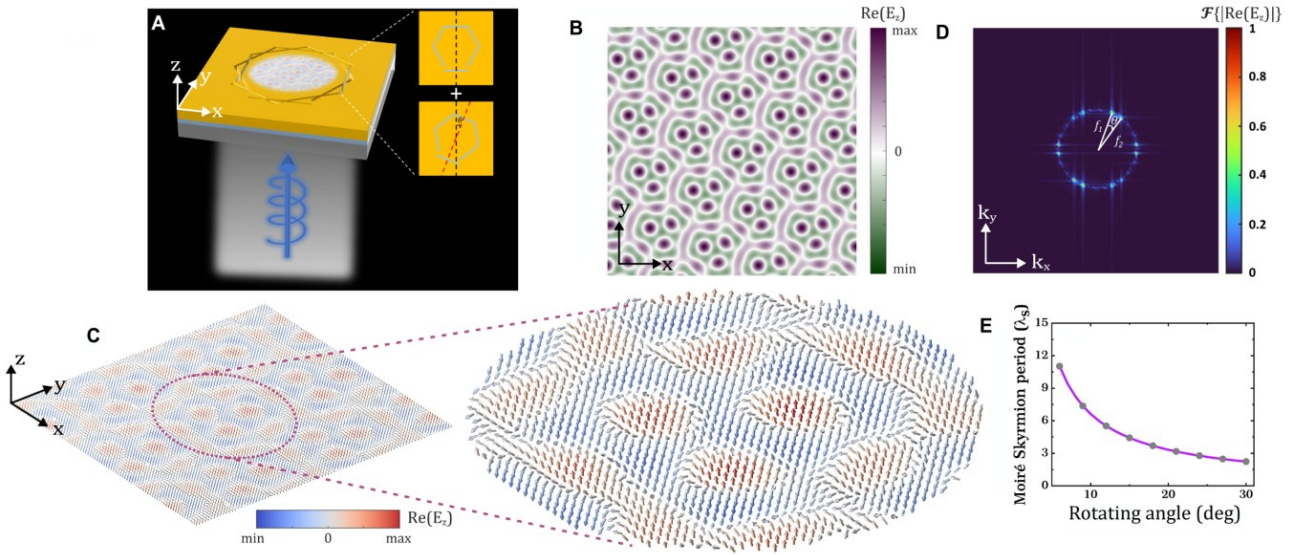

**Figure 1.** Moiré optical skyrmion cluster generator. (A) The nanostructure consists of two groups of hexagonal nanoslits with a relative twist etched on a gold layer, producing high-degree plasmonic skyrmion clusters. (B) Real component of the axial electric field at 38.21-degrees twist that sculptures the typical moiré

optical skyrmion superlattice. (C) Vector representation of the local unit electric field vector of the same twist angle of (B) showing optical skyrmion cluster. (D) Fourier spectral of the axial electric field component. (E) The relationship between twist angle and moiré optical skyrmion period.

To gain insight about the polarization topology, we calculate the skyrmion number $Q(\theta)$ per unit cell, which characterizes the number of "quasi-particle" skyrmion in a wave, is

$$Q(\theta) = \frac{1}{4\pi} \iint_{Cell} s(\mathbf{r}, \theta) d\mathbf{r}, \quad (3)$$

where the integral covers the unit cell, and s is the skyrmion number density (SND), takes the form of

$$s(\mathbf{r}, \theta) = \bar{\mathbf{E}} \cdot [\partial_x \bar{\mathbf{E}} \times \partial_y \bar{\mathbf{E}}]. \quad (4)$$

Deriving the transverse electric field from Maxwell's equations, we obtain the 3D polarization structure, as seen in Fig. 1C, and skyrmion number for each unit is calculated to be Q = -3. This is a manifestation of optical skyrmion cluster, which is in contrast to the elementary optical skyrmion (Q = ±1) lattice with $C_4$ and $C_6$ symmetry.

To interpret this, we consider interference of two skyrmion lattices $\mathbf{R}_a = n_1\mathbf{a_1} + n_2\mathbf{a_2}$ and $\mathbf{R}_b = n_1'\mathbf{a_1'} + n_2'\mathbf{a_2'} = \mathbb{R}\mathbf{R}_a$, where $n_1, n_2, n_1', n_2'$ are integers, and the lattice is described by two primitive lattice vectors $\mathbf{a_1} = (\alpha, 0)$ and $\mathbf{a_2} = (\alpha/2, \sqrt{3}\alpha/2)$ with $\alpha$ being the distance between the closest skyrmions. The interference of dual groups of skyrmion lattices described by $\mathbf{R}_a$ and $\mathbf{R}_b$ reshapes the topology of the electric field. In the case that $\mathbf{R}_a = \mathbf{R}_b$ holds for a set of integers $\{n_1, n_2, n_1', n_2'\}$, the resulting superimposed electric field remains periodic, while superimposed patterns in all other cases are quasi-periodic. The generated cell of moiré superlattice is much larger than the nontwisted case, with each cell contains multiple skyrmion units (See Supplemental Material S5). It is this trick that allows us to generate the optical skyrmion cluster that contains multiple skyrmion units, enabling not only the on-chip crystallization of optical skyrmion but also their quasi-crystallization.

To understand the plasmonic skyrmion clusters in momentum space, we proceed by analyzing the real component of the axial electric field in the spatial-frequency domain. This is achieved by performing a 2D Fourier transform of the axial field described by Eq. (1)

$$\tilde{S}(\mathbf{f}) = \sum_{j=1}^{N=12} \gamma_j \delta(\mathbf{k}_j - 2\pi \mathbf{f}), \quad (5)$$

where δ denotes the Dirac delta function, and the function $\gamma_j$ denotes the weight of the respective spatial frequency component. The vector $\mathbf{f}_j = \mathbf{k}_j/2\pi$ describes the frequency of the interfered SPP waves. This is manifested in the spatial frequency domain as interference of the lattice topology {$\mathbf{f}_1$, $\mathbf{f}_3$, ..., $\mathbf{f}_9$, $\mathbf{f}_{11}$} and the lattice topology {$\mathbf{f}_2$, $\mathbf{f}_4$, ..., $\mathbf{f}_{10}$, $\mathbf{f}_{12}$} with the twist angle $\theta$, and the magnitude $|\mathbf{f}_j|$ equals to $k_{spp}/2\pi$, see Fig. 1D.

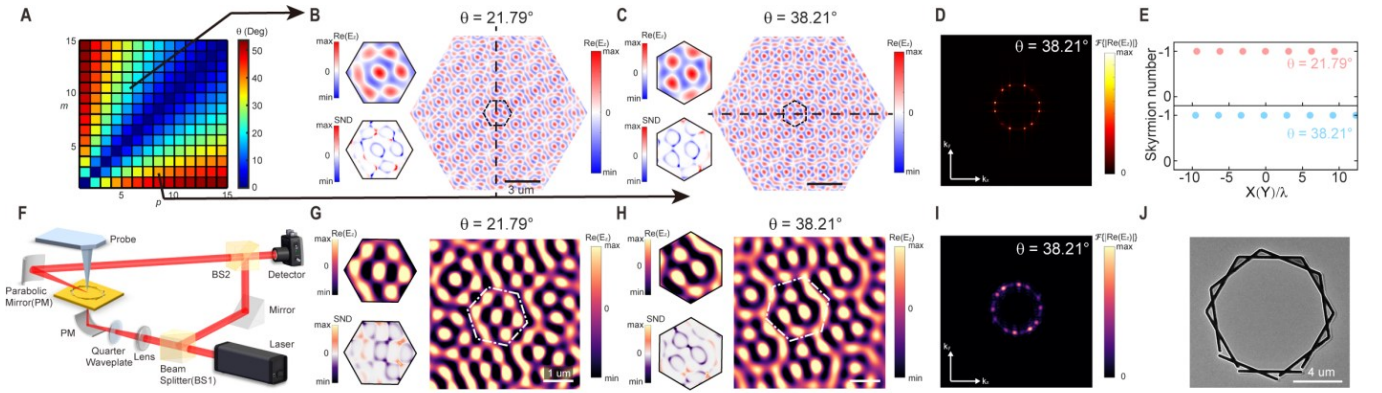

**Figure 2.** Simulation and experimental demonstration of the periodic plasmonic skyrmion clusters. (A) The commensurate twist angle for the generation of optical skyrmion superlattices. (B)-(C) Simulation results of moiré optical skyrmion superlattices with well-defined symmetry at several commensurate twist angles. (B) $\theta$ = 21.79°, (C) $\theta$ = 38.21°. Each figure contains simulated interference patterns of the real part of the axial field component. The unit cell of moiré optical skyrmions is marked with a black dotted line, the patterns of the real part (top) and skyrmion number density (SND) (bottom) are shown on the left. (D) Fourier spectral of the axial electric field component at 38.21 degrees. (E) The spatial distribution of skyrmions along a black dashed line at different twist angles. (F) Schematic of the experimental setup. (G)-(I) Experimental results counterpart of (B)-(D). (J) Scanning electron microscope image of a 21.79-degree twisted composite nanostructure sample.

## Experimental measurement of optical skyrmion clusters

To verify moiré optical skyrmion cluster predicted by this concept and obtain accurate optical responses, both full-wave numerical simulations and near-field experimental measurements were carried out. We fabricated samples with controlled angle to generate diverse skyrmion clusters, a scattering near-field scanning optical microscopy (Neaspec neaSNOM) was used to measure the amplitude and phase of the axial electric field on a 2D space via pseudo heterodyne interferometric, as illustrated in Fig. 2F (details in Supplemental Material). The Au film must have an adequate thickness to block light from areas without nanoslits, thereby avoiding unwanted waveguide phenomena and ensuring ease of fabrication. The nanoslits are carved in the 120 nm Au layer atopped a 1 mm glass substrate, using a 3-nm Ti adhesion layer between the gold and glass. The slits are 15 μm in length and 200 nm in width. The composite nanostructure is illuminated by a 633 nm laser with left-handed circular polarization at normal incidence, converting free-space propagating light to an SPP mode with moiré topology defects. The slit-scattering loss and slight position shifts of the nanoslits cause minimal amplitude asymmetry, which does not impair our results [32] (For the properties of excited surface waves, see Supplemental Materials).

**Periodic plasmonic skyrmion clusters**-The plasmonic skyrmion clusters are periodic only at specific twist angle, as the relation $R_b = R_a$ holds. For hexagonal lattices, this leads to the commensurate twists with $\cos\theta = [(p^2 + 4mp + m^2)/2(m^2 + mp + p^2)]$ parameterized by the integers $p$ and $m$ [33]. Figure 2A illustrates these angles for various indices $p$ and $m$. Depending on the value of $\theta$, the skyrmion number per unit cell changes.

SPPs are excited at the interface of air and gold in our system at two commensurate twist angles, i.e. 21.79 degrees and 38.21 degrees. Figure 2J shows the scanning electronic microscopic (SEM) image of the sample with a twist angle of 21.79 degrees. Figure 2G presents the measured axial electric field. Based on the experimental parameters, the simulated axial electric field and SND are shown in Figs. 2B and 2C. The spatial Fourier filtering technique is used to remove the noise and background to obtain the Fourier spectrum, as shown in Fig. 2I. This shows a good match to the numerical [Fig. 2D] and theoretical prediction [Fig. 1D]. The skyrmion cluster pattern can therefore be extracted. It can be clearly observed that each lattice unit cell contains one and three optical skyrmion units for 21.79 degrees and 38.21 degrees, respectively. They are periodically arranged in the 2D plane, which agrees with our theoretical expectations. This phenomenon is analogous to the recently reported skyrmion bags in liquid crystals [34]. Note that the twisting center is not perfectly located at the center of the skyrmion unit (See Supplemental Material S6). The deviation between simulation and experiment in plasmonic 3D polarization is attributed to the imperfect alignment, which includes wavevector and polarization of incident beam, as well as errors in sample fabrications. The change in skyrmion number as the twist angle varies is a result of the modulation of polarization, due to the interference between the two twisted polarization lattices.

The period of the plasmonic skyrmion cluster decreases sharply with the rotating angle $\theta$, viz., $g_m(\theta) = \alpha/\sqrt{2(1-\cos\theta)}, \theta \in (0, \pi/6)$ [35], and is related to the wavelength $\lambda_S$ $g_m(\theta, \lambda_S) = 2\lambda_S/\sqrt{6(1-\cos\theta)}$, see Fig. 1E. The optical skyrmion period at 21.79 degrees and 38.21 degrees are both $3.12\lambda_S$ [Fig. 2E], in accordance with the moiré period given by $g_m(\theta, \lambda_S)$. Compared to the conventional optical skyrmion lattice with a skyrmion number of one per lattice cell and fixed skyrmion density, higher-degree skyrmion cluster with skyrmion numbers larger than one can be formed.

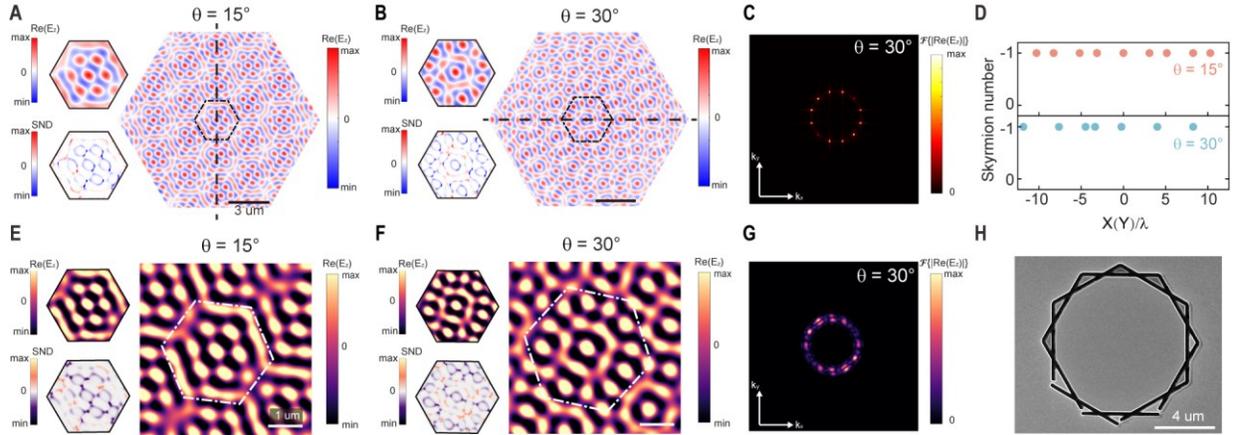

**Figure 3.** Simulation and experimental demonstration of the quais-periodic plasmonic skyrmion clusters. (A)-(B) Simulation results of moiré optical skyrmion superlattices with well-defined symmetry at several incommensurate twist angles. (A) $\theta$ = 15°, (B) $\theta$ = 30°. Each figure contains simulated interference patterns of the real part of the axial field component. The unit cell of moiré optical skyrmions is marked with a black dotted line, the patterns of the real part (top) and skyrmion number density (SND) (bottom) are shown on the left. (C) Fourier spectral of the axial electric field component at 30 degrees. (D) The spatial distribution of skyrmions along a black dashed line at different twist angles. (E)-(G) Experimental results counterpart of (A)-(C). (H) Scanning electron microscope image of a 30-degree twisted composite nanostructure sample.

**Quais-periodic plasmonic skyrmion clusters**-Of particular interest is the case as the twist angle is incommensurate. The SEM image of the sample with an incommensurate twist angle of 30 degrees is provided in Fig. 3H. The optical skyrmion lattices, referred to as super-hexagonal topology [36], can persist in the aperiodic case, as illustrated in Figs. 3A and 3E for a 15-degrees twist, containing five optical skyrmion units in the high-degree skyrmion cluster. Instead, for the 30-degrees twisted configuration [Fig. 3B and 3F], the optical skyrmion units form a more disordered pattern, lacking the symmetry observed in the 15-degree configuration. This disordered skyrmion arrangement can be identified in Fig. 3D. Compared to the numerical Fourier spectrum [Fig. 3C], the measurement [Fig. 3G] shows a double ring shape, which could be the unwanted phase ramp due to the synchronization during measurements. Moiré interference in composite nanostructures breaks the conventional periodic symmetry in skyrmions but unexpectedly spawns optical skyrmions topological states with quasi-crystalline order. This quasicrystallize state of optical skyrmion not only retains the topologically conserved properties of individual skyrmion (Q = -1), but also forms a Penrose-like lattice by quasi-periodic spatial arrangement.

Quasicrystallize states of optical skyrmion originate from the interference mismatch of the dual twisted skyrmion lattice, which reveals the coexistence of hidden long-range order and short-range disorder. It is noteworthy that the quasicrystallize state under 30 degrees twist has no periodic supercells. The optical skyrmion with multi-degree-of-freedom encoding capabilities (e.g., $Q$-value and $Q$-position) can empower the photonic integrated circuits for high-throughput optical information processing.

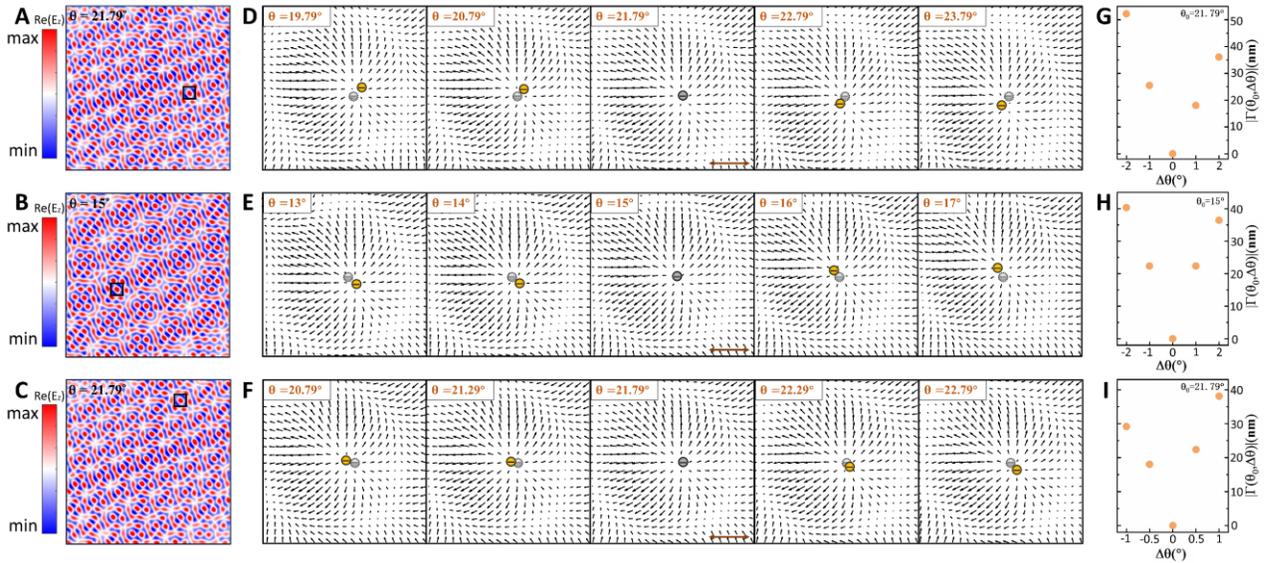

**Figure 4.** Nanoscale dynamics of optical skyrmions. (A)-(C) The real part of the axial electric field, spanning an area of 12 μm × 12 μm. The position of one skyrmion is selected in each image and marked with a black frame. (A) $\theta_0 = 21.79°$, (B) $\theta_0 = 15°$, (C) $\theta_0 = 21.79°$. (D)-(F) The movement of single skyrmion under small change in twist angles. The center of the skyrmion at each twist angle is marked by a yellow dot and is compared with the center of the skyrmion at twist angle $\theta_0$ marked by a gray dot, $\theta_0$ are (D) 21.79 degs, (E) 15 degs, (F) 21.79 degs. (D)-(F) correspond to the skyrmion marked in (A)-(C). (G)-(I) The relationship between $\Delta\theta = \theta-\theta_0$ and $|\Gamma(\theta_0,\Delta\theta)|$, here $|\Gamma(\theta_0,\Delta\theta)|$ is the distance between skyrmion center at angle $\theta$ and $\theta_0$. The scale bars are 200nm.

**Tracking the "quasiparticles" of light at nanoscale**
To further demonstrate the nanoscale behavior of plasmonic skyrmion clusters in our system, we study the configuration with small change of the twist angle, providing a platform to track the nanoscale change of optical skyrmion unit. Figures 4A-4C depict the out-of-plane electric field distribution in an area of 12 μm × 12 μm. Figures 4D–4F show the vector representation of the transverse electric field within the black boxes of Figs. 4A-4C. The center of optical skyrmion, determined by the location of the minimum transverse (in-plane) electric field, is pinpointed with its sign of skyrmion number.

In small change of the twist angles, the skyrmion topology is robust, but the nanoscale change in displacement of optical skyrmion is pronounced. In Figs. 4G-4I, we plot the skyrmion movement $|\Gamma(\theta_0,\Delta\theta)|$ as a function of the change in twist angle. The change of one degree induces the displacement of skyrmion around 20 nm. Such displacement of the skyrmion is not symmetric with positive and negative changes in the twist angle. This is expected since, the change of evanescent field is not linear with respect to the twist angle, but modulated by the cosine function, as can be seen from Eq. (1).

**Alignment deviation beacon via optical skyrmions**
Although the moiré optical skyrmion is robust to the small change of twist angle, the effect caused by the alignment deviation between the composite structures cannot be mitigated. We first consider non-twisted composite nanostructures $\theta = 0$, in which there is a lateral misaligned error $\Delta\mathbf{r}(\Delta x, \Delta y) = (x - x_s, y - y_s)$ [Fig. 5A]. Such polarization topology is distorted accompanied with symmetries breaking, the critical aspect of this misalignment lies in the emergence of topological transition. This takes place when two triangular skyrmion lattices interfere to generate a honeycomb lattice, thus resulting in the inversion of skyrmion number. The critical inversion condition of the optical skyrmion number is (see Supplemental Material S7)

$$\Delta\mathbf{r} = \frac{\varepsilon}{\sqrt{3}}(n_1\boldsymbol{a_1} + n_2\boldsymbol{a_2}); \quad (n_1 - n_2) \not\equiv 0 \,\{\text{mod}\,3\}, \quad (6)$$

where $\varepsilon$ is a 2x2 antisymmetric matrix that relates the coordinates of the composite structures and the optical skyrmion lattices. In the case that $(n_1 - n_2) \equiv 0\,\{\text{mod}\,3\}$, two sets of excited skyrmion lattices are coincides, showing the absence of optical skyrmion switching. To prove this, we numerically simulate the misaligned error and scan the skyrmion number per lattice versus the alignment deviations along the y-axis ($\Delta x = 0$, Fig. 5C) and the x-axis ($\Delta y = 0$, Fig. 5D), respectively. When the condition given by Eq. (6) is satisfied, rapid inversion of the optical skyrmion number from negative to positive is observed, and in the intermediate state, these optical skyrmions collapse. Note that inversion does not occur when the displacement is along the x axis [Fig. 5D]. The physical picture becomes trivial for the twisted cases $\theta \neq 0$, as the sudden switch of optical skyrmion state disappears due to the symmetry breaking of the superposed structures. Figures 5E-5G show the axial electric field for the structural alignment deviations $(\alpha, 0)$, $(\alpha/2, \sqrt{3}\alpha/2)$ and $(0, \sqrt{3}\alpha)$ at 30 degrees twist, respectively. The displacement induces the translation of the optical skyrmion superlattice, which can be interpreted from the change in the constant phase factor.

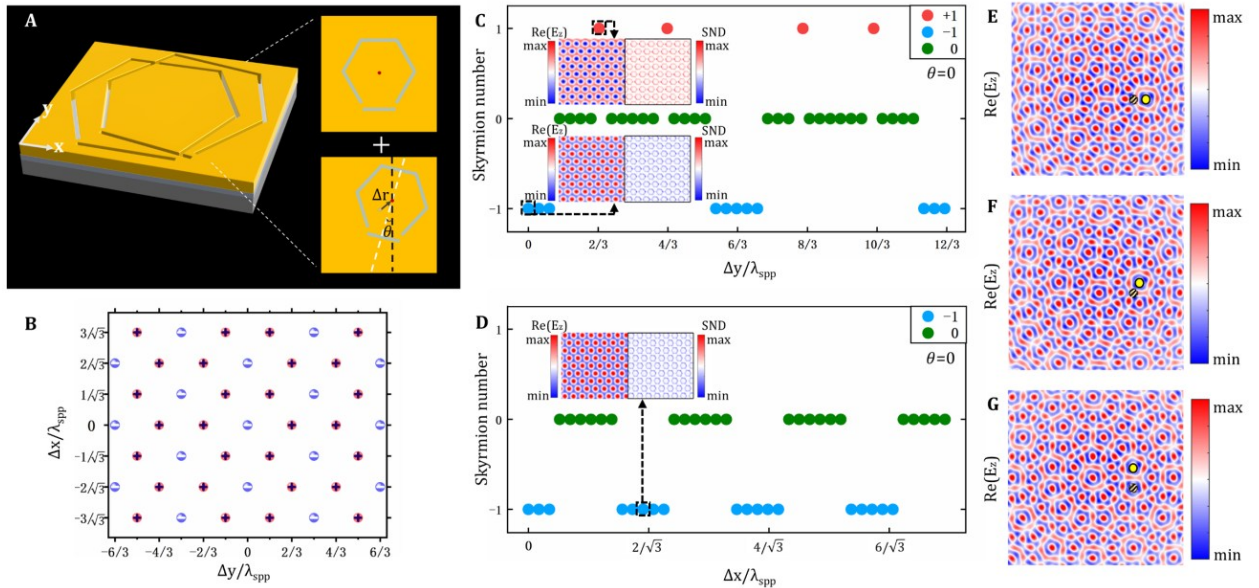

**Figure 5.** Optical skyrmions in the composite nanostructure with lateral misaligned error. (A) The structure consists of two sets of hexagonal slits with a relative displacement error $\Delta\mathbf{r} = (\Delta x, \Delta y)$. For the non-twist case, the theoretically predicted locations (B) where the skyrmion nucleate, with the red and blue circles representing the positive and negative skyrmion topological points. Optical skyrmion number along the y-axis (x = 0) (C) and x-axis (y = 0) (D), from which the skyrmion topology is observed to reverse along the y-direction. The orange, green and blue solid circles represent the positive, zero and negative skyrmion topology in the simulation scan. The insets present the real part of the axial electric field (left) and optical skyrmion number density (right) at $\Delta y/\lambda_{spp} = 0$ and 2/3 (C), and at $\Delta x/\lambda_{spp} = 2/\sqrt{3}$ (D). The electric field and skyrmion number density are plotted at 5*5 $\mu$m. For the twist case $\theta$ = 30 degs with displacement error. The real part of the axial electric field at (E) $\Delta x = \alpha, \Delta y = 0$; (F) $\Delta x = \alpha/2, \Delta y = \sqrt{3}\alpha/2$; (G) $\Delta x = 0, \Delta y = \sqrt{3}\alpha$. The line-filled circle in (F)-(G) marks an optical skyrmion unit when there is no alignment error, and the yellow solid circle represents the corresponding displaced skyrmion unit if the composite structure is misaligned.

## DISCUSSION

In conclusion, we theoretically propose and experimentally measure the plasmonic skyrmion clusters containing multi-skyrmions in a unit cell. Dual groups of optical skyrmion lattices are interfered to generate plasmonic skyrmion clusters via twistronics engineering of plasmonic nanostructures. This method provides a step towards high-degree optical skyrmions. The twist angle controls the nanoscale change of optical skyrmion, affording a powerful platform to engineer and explore the plasmonic skyrmion clusters in commensurate and incommensurate geometries. The experimental results corroborate our numerical and theoretical findings, which demonstrate the moiré-induced phenomena in twist-plasmonic systems. The formation of multi-skyrmion complexes showcases the potential of twist-angle degrees of freedom for controlling high-dimensional topology at subwavelength scales.

In the case of well-aligned composite structures, the resulting moiré pattern facilitates the formation of optical skyrmions with a well-defined skyrmion number. However, displacement errors of the structures can cause the switch of skyrmion topology, providing applications in the characterization of twisted bilayer structures and in the alignment of lithography mask. For instance, the nanometer-scale optical metrology could be achieved by measuring the change of skyrmion number. Our results are quite general, as it can not only be integrated into various photonic devices [37–39] and technologies [40–42], offering promising functionalities [43,44] and improved performance [45], but also be applied to other wave systems, ranging from water waves [46] to partially coherent waves [47–49]. More broadly, our results suggest that moiré engineering- previously confined to van der Waals materials - represents a universal strategy for manipulating quasi-particle topology across wave systems.

## MATERIALS AND METHODS
### Numerical Implementation
For numerical simulations of the near-field of the plasmonic system, a 3D finite-difference time-domain (FDTD) method is performed to solve Maxwell's equation numerically in time domain. The composite nanostructure and monitoring points are placed within a 35 μm × 30 μm × 1.5 μm volume, with perfectly matched layer (PML) boundaries are set in all the directions. A mesh with a resolution of 20 nm × 20 nm × 5 nm is used to cover both the nanostructure and the monitoring points, ensuring adequate sampling of the polarization state of the optical skyrmion. The thickness and permittivity of the gold film are set as 120 nm and $\varepsilon = -10.6 + 1.7i$ [50], respectively. In the simulation, a left circularly polarized Gaussian beam with a wavelength of 633 nm is incident on the plasmonic nanostructure. The beam waist is larger than the dimensions of the sample structure to ensure all the nanoslits are well excited.

### Sample Fabrication
The samples are fabricated on a 120 nm-thick Au layer, which was deposited previously on a 1 mm-thick glass substrate using Electron beam evaporation (HHV, AUTO500) with a deposited rate of 1 A/s. A 3-nm Ti is used as the adhesion layer between the gold and glass. Subsequently, the SPP coupling composite structures with nanoslits width 150 nm is etched using focused ion beam etching (Helios G4) from the metal side. This is controlled by the software named NanoBuilder. The etching parameters were set as 7 pA and 30 kV, respectively. The fabricated sample was verified by the scanning electron microscope (SEM), as shown in Supplementary Material S1. Recessed structures in these samples hinder the emission and detection of secondary electrons, leading to localized attenuation that appears as darker regions in the SEM images. The topological properties of skyrmions are highly sensitive to angular dislocations within composite nanostructures (see Supplementary Material S8), necessitating high precision in sample fabrication. The processing errors are minimized within acceptable limits. Specifically, for twist angles of 21.79°, 38.21°, 15°, and 30°, the processing errors are 1.3%, 0.8%, 1.1%, and 1.2%, respectively. These underscore the level of precision required to accurately capture the topological dynamics of skyrmions.

### Near-field Measurements
The amplitude of the near field signal is measured using a scattering (aperture-free) Near-field Scanning Optical Microscope (s-NSOM, Neaspec ltd. VIS-neaSCOPE+s type), with the experimental configuration schematically illustrated in Fig. 3A. A 633-nm continuous wave laser (Cobalt) is employed, with its beam split into two optical paths. One beam is weakly focused by a plano-convex lens, illuminating the sample from the glass side in a transmission mode configuration. The second beam is utilized for interferometric pseudo-heterodyne detection, which is modulated by a vibrating mirror with a frequency of 300 Hz interfered with the signal scattered from the sample to reconstruct the full electric field information, including amplitude and phase. To achieve circularly polarized incident light, a quarter wave plate is positioned in the optical path before the sample. The sample is positioned on a moving stage, where a silicon atomic force microscope (AFM) tip coated with platinum scattered predominantly the axial electric field component into a detector. The tip vibrates at oscillation frequencies between 200 and 230 kHz. The tapping amplitude is 50 to 70 nm. This nanotip scatters predominantly the axial electric field component into a detector through a parabolic mirror, where the integration time of detector is set to be 10-16 ms to ensure a good signal to noise ratio. The desirable electric field, stemming from the near field of the structure, is then demodulated at higher harmonics of the tip's vibration, to suppress the scattered free-space background signal [51]. In the measurements, the signal is extracted from the second demodulation order.

**Acknowledgements.** The authors would like to thank Neaspec GmbH and Quantum Design Co., Ltd. for help and guidance in experimental measurements and analysis.

**Funding**. National Natural Science Foundation of China (62305146, 12474299, 12164027); Training Program for Academic and Technical Leaders of Major Disciplines in Jiangxi Province (20243BCE51163); Natural Science Foundation of Jiangxi Province (20232BAB211031, 20242BAB20023) and Nanchang University Youth Training Program (PYQN20230064); the Project of Preeminent Youth Fund of Jiangxi Province (20224ACB211002); Jiangxi Provincial Key Laboratory of Photodetectors (No. 2024SSY03041).

**Competing interests:** the authors declare that they have no competing interests.

**Data and materials availability:** All data needed to evaluate the conclusions in the paper are present in the paper and/or the Supplementary Materials.

**Note Added**: After this work was completed, two related preprints, arXiv: 2411.00645 and arXiv 2411.03032, came to our attention.

**Supplementary Materials**
**This PDF file includes:**
Supplementary Text
Figs. S1 to S5